\documentclass[a4paper]{article}

\usepackage[utf8]{inputenc}
\usepackage[T1]{fontenc}
\usepackage{lmodern}

\usepackage{graphicx}
\usepackage{booktabs}
\usepackage{tabularx}
\usepackage{amsmath}
\usepackage{url}
\usepackage[margin=1.25in]{geometry}
\usepackage[sort&compress,numbers]{natbib}
\usepackage{titlesec}
\usepackage[font=small,labelfont=bf]{caption}
\usepackage[yyyymmdd]{datetime}

\titleformat*{\section}{\bfseries\large}
\titleformat*{\subsection}{\bfseries\normalsize}
\titleformat*{\subsubsection}{\bfseries\normalsize}

\newenvironment{csmtAbstract}{\noindent\small\bfseries Abstract:\ }{}
\newenvironment{keywords}{\small\noindent\bfseries Key Words:\ }{}

\newcommand*{\affaddr}[1]{{\sffamily\small (\textit{#1})}}
\newcommand*{\affmark}[1][*]{\textsuperscript{#1}}

\makeatletter
\renewcommand\@maketitle{%
  \begin{center}
    {\LARGE\bfseries \@title\par}
    \vskip 0.75em
    {\normalsize \@author\par}
  \end{center}\par\vskip 1em
}
\makeatother

\title{Oral Tradition-Encoded NanyinHGNN: Integrating Nanyin Music Preservation and Generation through a Pipa-Centric Dataset%
  \thanks{Acknowledgement: This work was supported by the Social Science Foundation of Fujian Province under Grant FJ2023JDZ050.}%
}

\author{%
  Jianbing Xiahou\affmark[1],\ Weixi Zhai\affmark[1],\ Xu Cui\affmark[2]\\
  \affaddr{\affmark[1]Quanzhou Normal University, Quanzhou, China}\\
  \affaddr{\affmark[2]School of Art, Xiamen University, Xiamen, China}
}

\begin{document}

	\maketitle
	
	\begin{csmtAbstract}
		We propose \textit{NanyinHGNN}, a heterogeneous graph network model for generating \textit{Nanyin} instrumental music. As a UNESCO-recognized intangible cultural heritage, \textit{Nanyin} follows a heterophonic tradition centered around the \textit{pipa}, where core melodies are notated in traditional notation while ornamentations are passed down orally, presenting challenges for both preservation and contemporary innovation.To address this, we leverage the characteristics of \textit{Nanyin} heterophony by constructing a \textit{Pipa-Centric} MIDI dataset, developing NanyinTok, a specialized tokenization, and converting symbolic sequences into graph structures using a Graph Converter, ensuring the preservation of key musical features.Our key innovation reformulates ornamentations generation as generating ornamentations nodes within a heterogeneous graph. First, a graph neural network generates melodic outlines optimized for ornamentations. Then, a Rule-Guided system, informed by \textit{Nanyin} performance practices, refines these melodies into complete ornamentations without requiring explicit ornamentation annotations in the training data. Experimental results demonstrate that our model successfully generates authentic heterophonic ensembles featuring four traditional instruments, thereby validating that the integration of domain-specific knowledge into model architecture effectively mitigates data scarcity challenges in computational ethnomusicology. 
        
        Our code is available at: https://github.com/wishzhai/NanyinHGNN
        \end{csmtAbstract}
	
	
	\section{Introduction}
	
	\textit{Nanyin}, inscribed on UNESCO’s Representative List of the Intangible Cultural Heritage of Humanity in 2009\footnote{https://ich.unesco.org/en/RL/nanyin-00199}, preserves a musical tradition that originated in China’s \textit{Minnan} region in the mid-10th century. As one of the most comprehensive surviving embodiments of ancient Chinese musical practices, spanning the \textit{Han}, \textit{Jin}, \textit{Tang}, and \textit{Song} dynasties. \textit{Nanyin} is often hailed as the "Living Fossil of Chinese Music History" due to its heterophonic texture. Compared to the complexity of Western music’s texture and harmony, \textit{Nanyin’s} heterophony is sometimes perceived as a relatively simple monophonic system\cite{Thrasher23Traditional}. In reality,  \textit{Nanyin’s} heterophony is an intricate and highly refined musical system. While performing the same fundamental melody, \textit{Nanyin} instrumentalists introduce variations based on the characteristics of their instruments. These variations follow a set of interactive principles that shape the musical texture:
\begin{itemize}  
\item \textbf{Skeletal Melodies:} \textit{Nanyin} music is built upon a fundamental shared melody recorded in \textit{GongQe Notation}. This core melodic framework, known as \textit{Skeletal Melodies}, serves as the backbone of composition, similar to figured bass in \textit{Baroque} music or lead sheets in \textit{jazz}. While providing essential pitch and rhythmic structure, it also allows for improvisational freedom.  

\item \textbf{Heterophony:} The \textit{Nanyin} ensemble consists of four core instruments: the \textit{Pipa} (pear-shaped lute), \textit{Sanxian} (three-stringed python-skinned lute), \textit{Dongxiao} (vertical bamboo flute), and \textit{Erxian} (two-stringed fiddle). The \textit{Pipa} and \textit{Sanxian} play the \textit{skeletal melodies} and techniques directly derived from \textit{GongQe Notation}, while the \textit{Dongxiao} and \textit{Erxian} introduce idiomatic ornamentations that enrich the musical texture.  

\item \textbf{Rule-Guided Improvisation:} While ornamentations in \textit{Nanyin} performance are improvised, they follow a set of interactive principles deeply rooted in oral tradition. These Rule-Guided improvisatory practices ensure stylistic consistency, preserving the essence of \textit{Nanyin} while allowing performers to introduce expressive nuances.  
\end{itemize}  

This duality of knowledge creates a preservation paradox: while static audio recordings document individual performances, they fail to capture the generative principles that sustain the dynamic practice of \textit{Nanyin}.

Our technical contributions unfold in three phases:

\begin{enumerate}
    \item \textbf{Dataset Creation.} We construct a \textit{Pipa-Centric} MIDI dataset for \textit{Nanyin} music, using a custom tokenization scheme, \textit{NanyinTok}, that captures essential performance features.
    
    \item \textbf{Graph Representation.} We develop a \textit{Structural Graph Converter} that transforms symbolic sequences into heterogeneous graphs, encoding melodic, rhythmic, and structural relationships.
    
    \item \textbf{Two-Stage Generation.} We design a two-stage system that generates core musical structure and applies rule-guided ornamentation to capture key \textit{Nanyin} techniques.
\end{enumerate}

Beyond its immediate applications to \textit{Nanyin} preservation and revitalization, our work contributes to broader discussions on the intersection of tradition and modernity in music information retrieval. By constructing a dataset that captures heterophony through its principal instrument and embedding domain knowledge into model design, we bridge symbolic representation and performance practice. This paves the way for future applications in education, restoration, and interactive composition, offering tools that support continuity in oral traditions. Our approach also addresses data scarcity by enabling constrained generation of stylistically grounded content, with future directions including human-in-the-loop systems aimed at preserving and revitalizing musical heritage.

	\section{Related Works}
\subsection{Music Heritage Revival}
The Bach Doodle project commemorates Johann Sebastian Bach's 334th birthday by leveraging machine learning to generate interactive musical compositions in his distinctive style. Utilizing the Coconet \cite{Liu24Coconet} the project masters Bach's complex contrapuntal structures through an innovative training approach of random score erasure and reconstruction. Despite the limited corpus of only 306 Bach compositions, this data augmentation technique dramatically expands the training dataset, enabling the model to accurately capture and reproduce Bach's sophisticated harmonic techniques and contrapuntal language.

In a different cultural context, 
Six Dragons Fly Again  \cite{Han24SixDragons}aims to revive the 15th-century Korean court music piece Flying Dragons, the existing version of which contains only a basic melody. This project employs a self-supervised BERT-like masked language model, trained on an OCR-recognized original score dataset, to fill in appropriate rhythms and notes, facilitating performance. It generates melodies played by the lead instrument, the \textit{Piri}, providing a foundation for ensemble generation. The system further optimizes the arrangement through collaborative multi-instrument coordination. Performance validation was achieved when the Korean National Gugak Center's Court Music Orchestra performed the generated results, completing a feedback loop from digital symbols to authentic cultural practice.
\subsection{Skeleton-Guided Generation}

Skeleton-guided melody generation integrates music-theoretical constraints through various approaches.  WuYun \cite{zhang2023wuyun} follows a two-stage process, extracting skeletons based on rhythm and pitch stability before refining them with neural networks. Small Tunes Transformer\cite{lv2025small} further incorporates hierarchical structures through phrase segmentation, skeleton note extraction, and Phrase-Level Cross-Attention, enhancing coherence with pentatonic constraints and Chinese-style trembling notes.

\subsection{Expressive Performance and GNN Modeling }

When it comes to ornamentations modeling and expressive performance, VirtuosoNet\cite{jeong2019virtuosonet} introduced a hierarchical RNN framework with attention mechanisms and measure-level estimation to generate expressive piano performances from MusicXML scores. Building on this foundation, the research team later developed a GNN model \cite{Jeong19GraphNN} that processes MusicXML data by capturing global structural relationships through graph-based representations. This GNN approach enables performances that align more closely with music-theoretical constraints.

Furthermore, advancements in graph neural networks have further enriched the field. GNNs show promise in polyphonic generation: Zou et al. \cite{Zou22Melons} model bar-level structures for long-term melody generation, while Cosenza et al. \cite{cosenza2023graph} disentangle rhythm and pitch via graph convolutions. Wu et al. \cite{wu2024graph} demonstrated GNN-based cross-modal alignment, refining instrumental arrangements to enhance coherence with vocal tracks

However, few works focus specifically on the heterogeneous nature of traditional music instruments and techniques. In contrast, our approach leverages the heterogeneous structure of traditional music, incorporating various instruments and performance techniques, to enhance the expressiveness and authenticity of generated compositions.

\section{Pipa-Centered Dataset }

Existing music datasets are heavily biased towards Western music, with non-Western genres accounting for only 5.7\% \cite{mehta-etal-2025-music}, limiting the cross-cultural capabilities of generative models. To address this gap and to preserve \textit{Nanyin} heritage, we introduce a new \textit{Pipa-Centric} paradigm rooted in computational ethnomusicology. Instead of relying on purely data-driven methods, we construct a fine-grained performance dataset and develop a novel approach that integrates domain knowledge. 
\subsection{From Score to Performance: A Practice Loop Grounded in Nanyin}

In \textit{Nanyin} ensemble music, the \textit{pipa} functions as both the principal melodic instrument and the temporal conductor. Through \textit{nianzhi} playing techniques, the \textit{pipa} initiates performances, guides tempo transitions, and signals closures with expressive \textit{ritardando}, directing the ensemble’s collective timing and phrasing. Supporting instruments such as the \textit{sanxian}, \textit{dongxiao}, and \textit{erxian} dynamically respond, forming \textit{Nanyin}'s characteristic heterophonic texture.

\begin{figure}[htbp]
    \centering
    \includegraphics[width=0.5\textwidth]{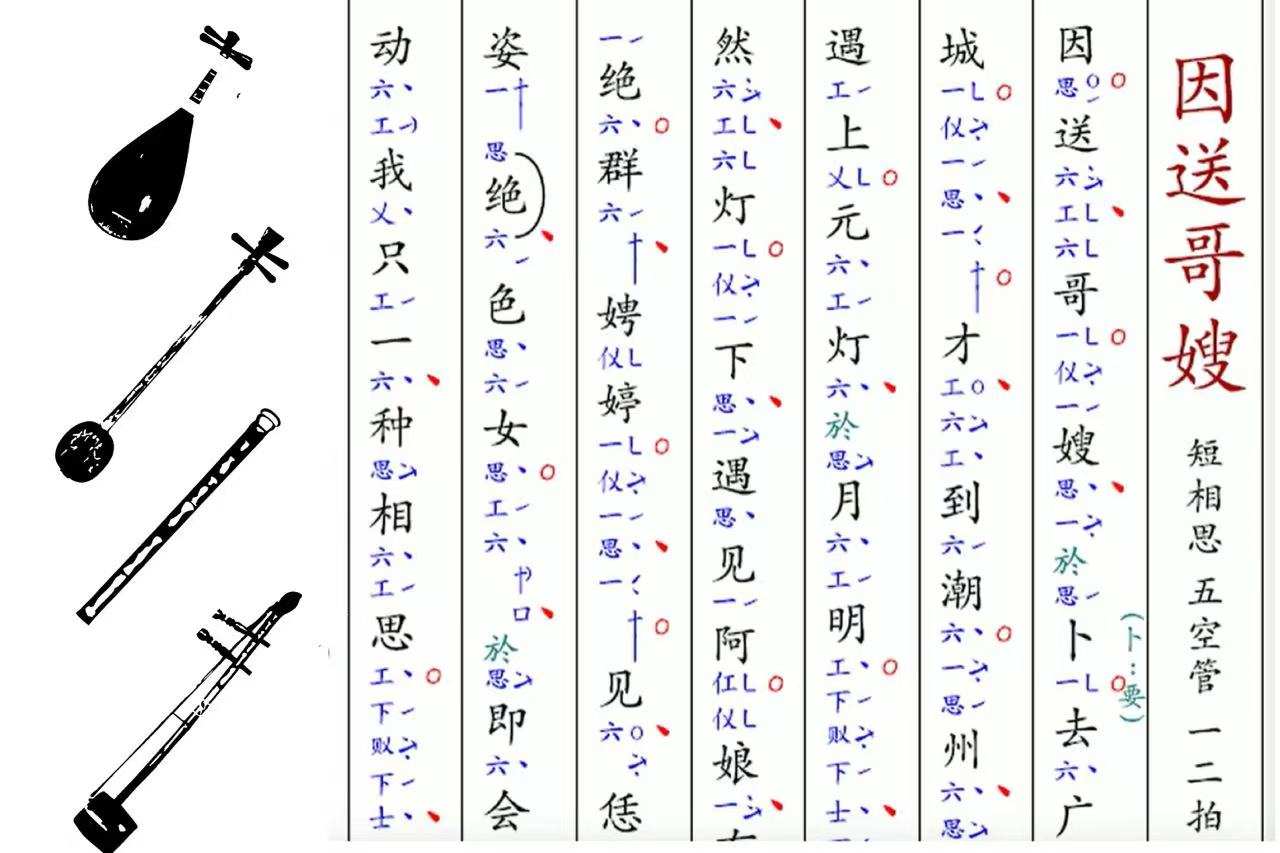} 
    \caption{Red indicates liaopai rhythm markers, blue represents pipa playing techniques and pitch notation, large black characters show the song lyrics, and green denotes chen-ci (filler words). }
    \label{fig:notation_graph}
\end{figure}

Traditional \textit{Nanyin} scores rely on \textit{GongQe} notation, a solmization-based pitch system where tones are defined by modal context and instrumental mapping, rather than absolute frequencies.
While culturally effective, this system omits rhythmic and expressive information, relying on oral transmission and embodied interpretation. A part of the the \textit{GongQe} notatis is illustrated in Figure~\ref{fig:notation_graph}.

To operationalize this performance logic computationally, We curated a culturally significant dataset comprising 332 \textit{Nanyin} \textit{pipa} compositions performed by conservatory-trained musicians specializing in this tradition. To achieve a reliable transcription of \textit{Nanyin} performance details, we employed Melodyne 5 for audio-to-MIDI conversion, followed by manual verification by trained musicians to ensure transcription accuracy.

       \subsection{NanyinTok: A Token System for Modeling Playing Technique}

To effectively model the expressive performance logic of \textit{Nanyin}, e developed NanyinTok, a specialized tokenization framework built upon MidiTok \cite{Fradet21MidiTok},
explicitly designed to encode not only pitch and timing, but also \textit{playing techniques} central to this tradition. Conventional MIDI event streams lack the ability to capture hierarchical phrasing, modal constraints, and gesture-driven ensemble interactions. 

\textbf{GongQe Notation Encoding.} We encode pitches based on \textit{GongQe} notation, mapping them to precise \textit{pipa} fingering locations within the modal system. The defined pitch set covers the full \textit{pipa} performance range:
\[
d,\ e,\ f,\ \#f,\ g,\ a,\ b,\ c^{1},\ \#c^{1},\ d^{1},\ e^{1},\ f^{1},\ \#f^{1},\ g^{1},\ a^{1},\ b\flat^{1},\ b^{1},\ c^{2},\ d^{2},\ e^{2},\ g^{2},\ a^{2},\ b^{2}
\]

The \textit{Nanyin} modal structure includes four pentatonic modes. For instance, the commonly used \textit{Wu-Kong} mode exhibits a bi-register organization: lower register pitches ($d$ to $a^{1}$) correspond to a C-based pentatonic scale, while upper register pitches ($d^{1}$ to $b^{2}$) form a G-based pentatonic system. To enforce modal fidelity during learning and generation, we apply dynamic masking, replacing out-of-mode tones with \textbf{UNK} tokens.

\textbf{Technique Tokens for Performance Modeling.} NanyinTok introduces specialized \textbf{Tech-Nianzhi} tokens marking occurrences of the \textit{nianzhi} gesture---a rapid, decrescendo plucking pattern central to \textit{pipa}'s conductor role. These gestures are detected via velocity and timing analysis, and categorized as fast, standard, or slow based on expressive variation.

In addition, NanyinTok encodes microtiming deviations and articulation density following PerTok principles~\cite{Lenz24PerTok}, capturing expressive fluctuations across instruments, such as breath control in the \textit{dongxiao} or textural dynamics in the \textit{sanxian}.

\section{Graph Converter}
We propose the Graph Converter framework to overcome a fundamental limitation in existing symbolic music tokenization. While powerful, current mainstream frameworks like MidiTok treat music as a linear sequence, ignoring the crucial internal structures and relational dependencies found in traditional forms like \textit{Nanyin}. Our framework, in contrast, uses a graph-based representation to effectively capture these non-sequential musical features.

\subsection{Heterogeneous Graph Structure}

GraphMuse \cite{Karystinaios2024GraphMuse} provides predefined graph representations for symbolic music analysis and classification but lacks the flexibility needed for generative tasks. In contrast, the Deep Graph Library (DGL) \cite{Wang19DeepGraphLibrary} enables heterogeneous graph modeling with deep learning integration, making it well-suited for capturing the heterophonic structures of \textit{Nanyin}. Therefore, we adopt DGL to construct a heterogeneous graph representation tailored to \textit{Nanyin}, ensuring flexibility for generative modeling.

\subsection{Graph-Based Rules Injection}

\textit{Nanyin's} ornamentations are primarily passed down through oral tradition. Interviews with \textit{Nanyin} artists and statistical analyses of academic studies reveal that adding major seconds above and below principal notes is the primary method for enriching \textit{Nanyin} vocalization. Notably, upper major second ornaments occur more frequently than lower ones. Meanwhile, \textit{diantiao} ornamentations follow a structured classification: \textit{Guan-style}, an anticipatory ornamentation applied before an upcoming \textit{nianzhi}, and \textit{Jie-style}, a post-note embellishment occurring when no \textit{nianzhi} follows.

\begin{figure}[htbp]
    \centering
    \includegraphics[width=1\textwidth]{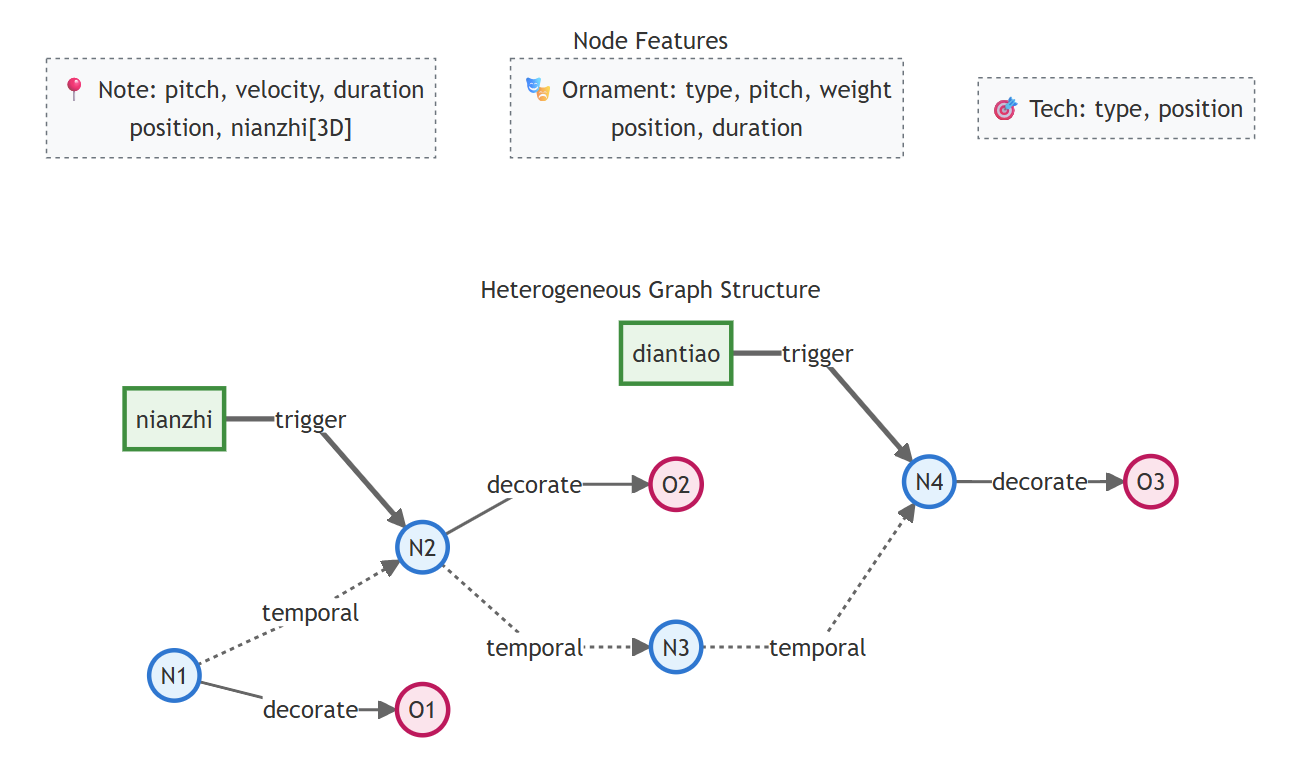} 
    \caption{Figure 1: The heterogeneous graph structure illustrates the relationships between musical notes (N), ornaments (O), and technical elements (green rectangles). Temporal connections are shown with dotted lines, decorative relationships with solid arrows, and trigger relationships with bold arrows. The nianzhi 3D feature vectors encode [is\_nianzhi, speed\_change, intensity\_pattern] to capture the complexity of this traditional Chinese playing technique.}
    \label{fig:heterogeneous_graph}
\end{figure}

We introduce a graph-based rule injection framework that embeds these \textit{Nanyin}-specific constraints into our heterogeneous graph representation. This is achieved through dynamic graph transformations that modify the graph's topology and features based on musical context, as illustrated in Figure~\ref{fig:heterogeneous_graph}.

The framework constructs the graph with three node types: \textbf{Note nodes}, which form the core musical sequence and contain features like pitch, velocity, and a specialized 3-dimensional \texttt{nianzhi} vector \texttt{[is\_nianzhi, speed, intensity]}; \textbf{Ornament nodes}, which are dynamically added to capture decorative elements with features such as type and weight; and \textbf{Tech nodes}, which represent performance techniques. The rule injection mechanism then applies three main enhancements: \textbf{Pentatonic Enhancement}, which boosts the weights of decorative edges that follow five-tone scale progressions by a factor of 2.0; \textbf{Intelligent Ornament Placement}, which dynamically adds ornament nodes to achieve a target density of 0.6 based on melodic context and probabilistic interval selection (favoring upper major seconds); and \textbf{Technique-Aware Processing}, which models the impact of techniques like \textit{nianzhi} and \textit{diantiao} by creating trigger edges and further modifying the graph. This dynamic graph transformation approach ensures that our model learns not just a musical sequence, but the underlying cultural rules that govern authentic \textit{Nanyin} performance.

 \section{Two-Stage Training Framework}

We introduce a two-stage training framework designed for \textit{Nanyin} generation. This framework consists of two distinct phases: skeletal melody generation and ornamentation system modeling, each focusing on different hierarchical aspects of \textit{Nanyin's} melodic and ornamental structure.

\subsection{Stage 1: Skeletal Melody Generation with Enhanced GATv2}

In this stage, we propose an enhanced GATv2\cite{BrodyAttentive} -based graph neural network for \textit{Nanyin} melody generation. Our model employs a 3-layer GATv2 architecture with hidden dimension of 256 and 4 attention heads, trained from scratch specifically for \textit{Nanyin} music generation. The model incorporates a feature enhancer with multi-scale processing capabilities and an enhanced pitch predictor with improved attention mechanisms, balancing efficiency and expressivity to provide a strong foundation for ornamentations.

The loss function for Stage 1 combines multiple component:

\begin{equation}
\mathcal{L}_{stage1} = \mathcal{L}_{CE} + \lambda_1 \mathcal{L}_{L1} + \lambda_2 \mathcal{L}_{consistency}
\end{equation}

where $\mathcal{L}_{CE}$ is the categorical cross-entropy loss for pitch prediction with class weighting, $\mathcal{L}_{L1}$ is the L1 regularization loss with $\lambda_1 = 0.0001$, and $\mathcal{L}_{consistency}$ is the feature consistency loss measuring Euclidean distances to class centers with $\lambda_2 = 0.1$.

Training is conducted with learning rate 0.0003, batch size 8, gradient clipping threshold 1.0, and cosine annealing scheduler ($T_0=20$, $T_{mult}=2$, $\eta_{min}=0.00002$). Early stopping monitors validation loss with patience of 35 epochs and minimum improvement threshold of 0.0003. This targeted approach specifically improves \textbf{pitch prediction} (adapting to the pentatonic scale), \textbf{duration modeling} (capturing \textit{Nanyin's} rhythmic structures), and \textbf{dynamic control} (ensuring expressive variations).

\subsection{Stage 2: Self-supervised Learning for Nianzhi Prediction and Rule-Guided Ornamentation}

In this stage, we develop a self-supervised learning framework for \textit{nianzhi} prediction and a rule-based ornamentation system, both designed to capture the stylistic nuances of \textit{Nanyin}. Given the absence of explicit ornamentation annotations in our data set, we reframe ornamentations generation as a \textbf{node creation task within a heterogeneous graph structure}, integrating computational ethnomusicology with oral-traditional performance heuristics.

\subsubsection{Nianzhi Prediction}

\textit{Nianzhi} is a distinctive plucking technique in \textit{Nanyin}, characterized by gradual changes in intensity and speed within a repeated note. To model this expressive variation, we introduce a self-supervised learning module that learns to predict when and how \textit{nianzhi} should be applied based on musical context. A deep feature extractor transforms input embeddings into a compact feature space, while a multi-head attention mechanism captures contextual dependencies crucial for \textit{nianzhi} placement. The model outputs three key parameters: position probability (threshold 0.7), speed variation (3-4 repetitions), and intensity pattern (pitch range 55-85).

To ensure accurate prediction, we optimize the following objective:
\begin{equation}
\mathcal{L}_{\text{nianzhi}} = \alpha_p \mathcal{L}_{\text{position}} + \alpha_s \mathcal{L}_{\text{speed}} + \alpha_i \mathcal{L}_{\text{intensity}}
\end{equation}

where $\mathcal{L}_{\text{position}}$ ensures correct placement with $\alpha_p = 0.35$, $\mathcal{L}_{\text{speed}}$ captures speed variations with $\alpha_s = 0.25$, and $\mathcal{L}_{\text{intensity}}$ models intensity patterns with $\alpha_i = 0.15$.

\subsubsection{Graph-Based Ornamentation Generation}

Building upon the \textit{Skeletal Melodies} generated in the first stage, we formalize ornamentation as a structured decision-making process within a heterogeneous graph representation of \textit{Nanyin}.

Inspired by \textit{Nanyin} performance traditions, we define three primary types of ornaments: \textbf{Standard Ornaments} (duration factor 0.3, velocity factor 0.9), \textbf{Light Appoggiaturas} (duration factor 0.2, velocity factor 0.8), and \textbf{Melodic Integration Ornaments} (duration factor 0.3, velocity factor 0.9). Each type is governed by specific \textbf{duration factors, velocity factors, and pitch intervals}.

For each selected position, the system generates appropriate ornamentations using the following mechanisms:
\begin{itemize}
    \item \textbf{Pitch Selection}: The system prioritizes major second intervals (+2) with 90\% weight and minor third intervals (+3) with 10\% weight, maintaining \textit{Nanyin's} characteristic melodic contours.
    \item \textbf{Positioning}: Grace notes are placed 0.015 time units before the main note, while after-notes are positioned 0.015 time units after the main note, ensuring natural musical flow.
    \item \textbf{Stylistic Adaptation}: The \textit{OrnamentProcessor} employs contextual analysis with a 6-note window to fine-tune ornamentation parameters based on three predefined stylistic modes.
\end{itemize}

The ornamentation parameters are controlled by style-specific rules with global constraints including ornament probability of 0.4, minimum interval of 3 time units between ornaments, and evaluation metrics targeting density (0.2-0.6), coverage (0.7), and evenness (0.8). This parametric design allows the model to generate diverse ornamentation patterns that align with \textit{Nanyin} traditions while maintaining musical coherence.

\section{Experiments}

\subsection{Dataset and Training}

We manually divided 337 \text{Nanyin} MIDI tracks into 507 segments, each not exceeding three minutes, ensuring that the integrity of the \textit{nianzhi} and other musical semantics were preserved during segmentation. The dataset was first randomly shuffled, followed by a split into training and validation sets, with 432 allocated for training. The decoding process aimed to generate multi-track MIDI from graph-structured input data, using a custom \text{Nanyin} decoder designed specifically for this task. This decoder was able to handle multiple instrument tracks, including the main \text{pipa}, as well as the \textit{sanxian}, \textit{dongxiao}, and \text{erxian}, each with its own set of MIDI generation rules. The training was conducted using a single Nvidia RTX 4070 GPU. 

\subsection{Evaluation Metrics}
We use a weighted F1-score to evaluate the model’s pitch prediction accuracy, with the weights determined by the frequency of each pitch class in the training dataset. To better reflect the modal characteristics of \textit{Nanyin} music, we also compute a mode-aware F1-score by focusing on the pitch classes within the \textit{Nanyin} modal system. This ensures a more accurate assessment of the model’s ability to preserve modal consistency.

For evaluating ornament quality, we introduce the Ornament Rationality Score (ORS), which assesses both stylistic coherence and structural distribution. Stylistic rationality measures how well the ornaments integrate with the melody in terms of melodic coherence and rhythmic alignment, while structural rationality quantifies the appropriateness of ornament distribution in terms of density, evenness, and coverage. The final ORS is a weighted sum of these components, offering a comprehensive indicator of ornamentation quality in the generated \textit{Nanyin} music.

For the \textit{Nianzhi}-specific evaluation, we adopt the loss functions from Section 5.2.1 to ensure consistency in evaluating the model’s ability to generate accurate and stylistically appropriate \textit{nianzhi}.

\section{Inference}
To generate a complete 	\textit{Nanyin} ensemble, we employ a structured pipeline that integrates learned models with rule-based refinements, ensuring stylistic coherence and authenticity.
\subsection{Ensemble Generation} 
 
For \textit{nianzhi} generation, we use a three-step process: a bi-directional LSTM detects potential positions based on multi-modal features, a note expansion mechanism enforces pitch consistency while applying exponential decay to velocity and duration, and a self-supervised module refines placement using dynamic thresholding.  

For ornamentation generation, we implement a hierarchical rule-based system. Structural rules ensure pentatonic-scale prioritization and modal consistency. Pattern-based rules employ an LSTM-driven embedding mechanism to model idiomatic \textit{Nanyin} gestures. Performance rules apply instrument-specific modifications: \textit{sanxian} notes shorten by 20\%, \textit{dongxiao} transposes down an octave, and \textit{erxian} shifts up by 12 semitones. The final ornamentation density is sampled from a normal distribution centered at 0.6.  

Melodies for all four instruments (\textit{pipa, sanxian, dongxiao, erxian}) are generated sequentially, with each instrument’s output conditioning the next. A final refinement step regenerates each melody while incorporating the full ensemble context, reinforcing \textit{Nanyin}’s interdependent heterophonic texture.  
\subsection{Special Note Seed}
Nanyin is characterized by the distinctive use of C\# and F\#, which play a crucial role in shaping its unique tonal flavor. Classic Nanyin pieces such as \textit{Shan Xianjun}, \textit{Yuanxiao Shiwuri}, and \textit{Wo Wei Ni} prominently showcase the expressive qualities of these notes within Nanyin’s modal system. While Western music also utilizes C\# and F\#, Nanyin’s interpretation of these tones produces a distinctively native aesthetic. This is because these notes often form non-native pitch combinations, such as $A$–$C\sharp$ and $D$–$F\sharp$, introducing brief tonal modulations that momentarily diverge from the traditional pentatonic structure. This technique serves as a key contributor to Nanyin’s distinctive melodic fingerprint. To ensure that the generated MIDI exhibits the characteristic tonal qualities of \textit{Nanyin}, we propose a \textbf{Special Note Seed} ornamentation generation mechanism. The Special Note Processor analyzes local tonal context, interval relationships, and pitch trajectories to compute a suitability score for each candidate position. It then selects the most appropriate special note as the ornamentation pitch while inheriting the timing of the main note, thereby generating an ornamentation sequence that aligns with the stylistic features of \textit{Nanyin}.

The model employs a temperature-controlled sampling mechanism (temperature = 0.8) to balance between creativity and adherence to traditional patterns. This approach ensures that while the generated melodies maintain the distinctive use of C\# and F\#, they do so in a way that preserves the natural flow and authenticity of Nanyin music. 

\subsection{Expert Reviews} 
To validate our approach, we rendered the generated MIDI using traditional Nanyin instruments. Grounded in Nanyin authenticity, our evaluation framework assessed the expressiveness of nianzhi, the coordination of four-instrument ensembles, the naturalness of ornamentations, and the preservation of Nanyin aesthetic nuances, particularly in genre-specific melodic contour and phrasing. Five expert Nanyin performers and researchers participated in a single-blind evaluation, 

Participants rated both samples on a 5-point Likert scale (1=low, 5=high) across the four criteria. Subjective feedback further contextualized quantitative findings. Results are summarized in the table below:
The results are summarized in Table~\ref{tab:Expert Reviews}.

\begin{table}[!ht]
\centering
\small
\label{tab:eval_results}
\begin{tabularx}{0.90\linewidth}{>{\raggedright\arraybackslash}X >{\centering\arraybackslash}X >{\centering\arraybackslash}X}
\toprule
\textbf{Criterion} & \textbf{With Seeds} & \textbf{No Seeds} \\
\midrule
\textit{nianzhi}  & 4.7 (±0.3) & 3.2 (±0.5) \\
Coordination  & 4.5 (±0.4) & 3.4 (±0.6) \\
Ornamentations & 4.3 (±0.4) & 2.9 (±0.7) \\
Aesthetic & 4.6 (±0.3) & 3.1 (±0.5) \\
\bottomrule
\end{tabularx}
\caption{Expert Evaluation Results (Mean ± Standard Deviation)}
\vspace{6pt}
\fontsize{10}{12}\selectfont
\label{tab:Expert Reviews}
\end{table}

This validates our methodology's effectiveness in preserving Nanyin's cultural acoustics while providing computational framework for heritage music revitalization.

\section{Conclusion}\label{sec:conclusion}

This work establishes a new paradigm for computational modeling of Nanyin music through heterogeneous graph neural networks, addressing the fundamental tension between prescriptive notation and improvisational practice for music heritage.

	\bibliographystyle{gbt-7714-2015-numerical}
	\bibliography{main}
	
\end{document}